\documentclass[10pt,a4paper]{article}
\usepackage{amsfonts}
\usepackage{pdfsync}
\usepackage{amsmath}
\usepackage{amssymb}
\usepackage{amsthm}
\usepackage{mathrsfs}
\usepackage{latexsym}
\usepackage[utf8]{inputenc}
\usepackage[T1]{fontenc}
\newtheorem{theorem}{Theorem}[section]
\newtheorem{lemma}[theorem]{Lemma}
\newtheorem{corollary}[theorem]{Corollary}
\newtheorem{proposition}[theorem]{Proposition}

\theoremstyle{definition}
\newtheorem{remark}[theorem]{Remark}

\newcommand{\dd}{\mathrm{d}}
\newcommand{\R}{{\mathord{\mathbb R}}}

\newcommand{\N}{{\mathord{\mathbb N}}}

\newcommand{\dom}[1]{\mathrm{Dom}(#1)}

\newcommand{\ee}[1]{\mathrm{e}^{#1}}
\newcommand{\ie}{\emph{i.e.}}

\newcommand{\cf}{\emph{cf}}
\newcommand{\grad}{\mathop{\mathrm{grad}}\nolimits}
\newcommand{\divergence}{\mathop{\mathrm{div}}\nolimits}
\newcommand{\curl}{\mathop{\mathrm{curl}}\nolimits}

\newcommand{\eps}{\varepsilon}
\usepackage{color}

\begin{document}
\title{\bf The magnetic Laplacian in shrinking 
tubular neighbourhoods of hypersurfaces}
\author{
D. Krej\v{c}i\v{r}\'{i}k,$^{a}$ 
N. Raymond$^{b}$ and 
M. Tu\v{s}ek$^{c}$}
\date{
\small
\emph{
\begin{quote}
\begin{itemize}
\item[$a)$]
Department of Theoretical Physics,
Nuclear Physics Institute ASCR, 25068 \v{R}e\v{z}, Czech Republic;
krejcirik@ujf.cas.cz
\item[$b)$]
Institut de Recherche Math\'ematiques de Rennes,
Universit\'e de Rennes~1, 
35042 Rennes Cedex, France;
nicolas.raymond@univ-rennes1.fr
\item[$c)$]
Department of Mathematics,
Faculty of Nuclear Sciences and Physical Engineering,
Czech Technical University in Prague,
Trojanova 13, 12000 Prague 2, Czech Republic;
tusekmat@fjfi.cvut.cz
\end{itemize}
\end{quote}
}
\medskip
18 March 2013
} 
\maketitle
\begin{abstract}
\noindent
The Dirichlet Laplacian between two parallel hypersurfaces
in Euclidean spaces of any dimension 
in the presence of a magnetic field
is considered in the limit when the distance between
the hypersurfaces tends to zero.
We show that the Laplacian converges in a norm-resolvent sense
to a Schr\"odinger operator on the limiting hypersurface 
whose electromagnetic potential is expressed in terms of 
principal curvatures and the projection of the ambient 
vector potential to the hypersurface.
As an application, we obtain an effective approximation
of bound-state energies and eigenfunctions in thin quantum layers.
\end{abstract}

\newpage
\section{Introduction}
%
Given a hypersurface~$\Sigma$ in the Euclidean space~$\R^d$ with $d \geq 2$,
consider a charged quantum particle constrained to 
a tubular neighbourhood
\begin{equation}\label{layer.intro}
  \Omega_\eps := \big\{x+t\,n \in \R^d \ \big| \ 
  (x,t) \in \Sigma \times (-\eps,\eps) \big\}
  \,,
\end{equation}
where~$n$ denotes a unit normal vector field of~$\Sigma$.	 
This paper is inspired by the following questions:
What is the effective dynamics on the hypersurface
approximating the constrained motion when $\varepsilon \to 0$?
How does the former depend on the geometry of~$\Sigma$? 
To what extent can one recover the geometry of~$\Sigma$
from the energy spectrum?
What is the role of dimension~$d$?
What is the effect of interaction 
with an ambient magnetic field $B=*\dd A$?

In this paper we model 
the constrained quantum Hamiltonian 
by the magnetic Laplacian 
\begin{equation}\label{Laplacian.intro}
  (-i\nabla+A)^2 
  \qquad \mbox{on} \qquad 
  L^2(\Omega_\eps)
  \,,
\end{equation}
subject to Dirichlet boundary conditions on~$\partial\Omega_\eps$,
and tackle the questions by developing a singular perturbation theory
for a self-adjoint realization of the operator
that we concisely denote by $-\Delta_{D,A}^{\Omega_\eps}$.
Our main result says that 
\begin{equation}\label{main.intro}
  -\Delta_{D,A}^{\Omega_\eps} - \left(\frac{\pi}{2\eps}\right)^2
  \ \xrightarrow[\eps \to 0]{} \
  h_\mathrm{eff} := -\Delta_{D,A_\mathrm{eff}}^{\Sigma}
  + V_\mathrm{eff}
\end{equation}
in \emph{a norm-resolvent sense} under some additional assumptions on $\Sigma$ and $A$ 
(see Theorem \ref{Thm.main.bis} for the precise formulation).
Here the kinetic part of the effective Hamiltonian~$h_\mathrm{eff}$
is the magnetic Laplace-Beltrami operator on~$\Sigma$,
subject to Dirichlet boundary conditions on~$\partial\Sigma$ 
if the hypersurface is not complete,  
with~$A_\mathrm{eff}$ being just the projection of~$A$ on~$\Sigma$.
The \emph{geometric potential}~$V_\mathrm{eff}$ depends explicitly on  
principal curvatures of~$\Sigma$, \cf~\eqref{V.eff}.
The subtraction of the diverging coefficient in~\eqref{main.intro}
is needed in order to ``filter out'' 
the transverse oscillations due to the approaching Dirichlet boundaries.
It follows that the limiting operator~$h_\mathrm{eff}$ 
and its spectrum contain information about 
both the intrinsic and extrinsic geometry of the hypersurface~$\Sigma$.
The role of~$A_\mathrm{eff}$ is best visualized for $d=3$
where $B_\mathrm{eff} := \curl A_\mathrm{eff} = n \cdot B$,
\ie~only the projection of the ambient magnetic field to the normal 
bundle of~$\Sigma$ plays a role in the limit.

Except for the inclusion of magnetic field --
which is a primary motivation for us to write this paper --
the aforementioned questions have been considered by several authors
during the last two decades,
in various settings and with different methods.
Indeed, we dare to say that 
the schematic result~\eqref{main.intro} for $A=0$,
especially the presence of the geometric potential~$V_\mathrm{eff}$
in the limit,  
belongs now to an almost common knowledge among 
spectral geometers and mathematical physicists. 
Within the range of numerous papers on the subject,
let us point out just a few contributions,
closest to the present setting,
\cite{JK,daC1,daC2,Tol,DE,Mitchell,FrHe,
FK4,Wittich_2008,
Wachsmuth-Teufel,Wachsmuth-Teufel-short,Lampart-Teufel-Wachsmuth,
KSed}.
 
Despite of the extensive literature, 
there does not seem to be any work dealing with~\eqref{main.intro}
in the complete setting of the present paper.
More specifically, we are not aware 
of any work establishing~\eqref{main.intro}
through the norm-resolvent convergence,
for arbitrary hypersurfaces (bounded or unbounded),
in any dimension and notably with the presence of magnetic field. 
Furthermore, in this paper we propose a remarkably simple
method how to establish the norm-resolvent convergence
as a consequence of certain operator inequalities,
which we believe is of independent interest.
Finally, our technique provides an explicit bound on the decay 
rate of the limit~\eqref{main.intro}.

For $d=2,3$ 
the result~\eqref{main.intro} has applications 
in mesoscopic physics~\cite{LCM,Hurt}.
It can be used to approximate two- or three-dimensional
quantum dynamics in long thin nanostructures by
the one- or two-dimensional effective Hamiltonian, respectively.
Probably the most spectacular phenomenon here is 
the existence of curvature-induced bound states 
in unbounded
\emph{quantum waveguides} ($d=2$) \cite{ES,DE,KKriz}
and \emph{quantum layers} 
($d=3$) \cite{DEK2,CEK,LL1,LL2,LL3,Lu-Rowlett_2012}.
The purely quantum effect can be well understood from~\eqref{main.intro}:
since the geometric potential~$V_\mathrm{eff}$ 
is always non-positive for $d=2,3$, \cf~\eqref{V.eff},
it represents an \emph{attractive} interaction
and therefore generates discrete eigenvalues below the essential
spectrum of~$-\Delta_{D,0}^{\Omega_\eps}$.

Our new result involving the magnetic field gives a recipe 
how to possibly eliminate the disturbing bound states:
just embed the device in an ambient magnetic field 
and employ the \emph{repulsive} (diamagnetic) 
feature of the latter \cite{Laptev-Weidl_1999}.
For $d=2$ and $\eps$~\emph{fixed}, the repulsive nature 
of the magnetic field in the waveguide context
was previously studied in~\cite{MK-Kov}.
In higher dimensions ($d \geq 4$), 
no robust results about
the existence of discrete eigenvalues can be expected,
which can be again seen from~\eqref{main.intro}: 
in addition to different threshold properties of $-\Delta_{D,0}^\Sigma$,
$V_\mathrm{eff}$ may be repulsive.	

The paper is organized as follows. 
In the forthcoming Section~\ref{Sec.Pre}
we recall elements of geometry of hypersurfaces
and introduce a natural parametrization of~$\Omega_\eps$. 
Basic information about the magnetic field
in arbitrary dimension and curvilinear coordinates
are summarized in Section~\ref{Sec.field}.
In Section~\ref{Sec.Laplace} we introduce 
the magnetic Laplacian $-\Delta_{D,A}^{\Omega_\eps}$
and a unitarily equivalent operator~$H$ on 
an $\eps$-independent Hilbert space,
in terms of which the convergence result~\eqref{main.intro}
will be stated.
Section~\ref{Sec.compar} is devoted to various 
two-sided estimates of~$H$;
here the geometric potential~$V_\mathrm{eff}$
and the effective Hamiltonian~$h_\mathrm{eff}$
are encountered for the first time
in the present analysis.
The main idea how to deduce the norm-resolvent convergence
from the estimates is contained in Section~\ref{Sec.nrs},
where we also establish the main result of this paper
(Theorem~\ref{Thm.main}).
An alternative version of the result (Theorem~\ref{Thm.main.bis}), 
which is more closer 
in its spirit to the schematic limit~\eqref{main.intro},
is proved in Section~\ref{Sec.reduction}
by means of an orthogonal decomposition of the Hilbert space.
In Subsection~\ref{Sec.phys} we make the general results 
more explicit for the physically interesting situations of $d=2,3$.
Finally, in Section~\ref{Sec.spec} we show how the theorems
can be used to deduce a convergence of eigenvalues and eigenfunctions
(Corollary~\ref{Thm.spec}). 

\section{Geometric preliminaries}\label{Sec.Pre}
%
Let~$\Sigma$ be a connected orientable $C^3$ hypersurface 
(compact or non-compact) in~$\R^d$, with $d \geq 2$,
equipped with the Riemannian metric~$g$ induced by the embedding. 
The orientation is specified by a globally defined 
unit normal vector field $n:\Sigma\to S^{d-1}$.
Without loss of generality, we assume that~$\Sigma$
has the same orientation as the ambient Euclidean space~$\R^d$. 

For any $x\in\Sigma$, we introduce the \emph{Weingarten map} 
$$
  L: T_x\Sigma \to T_x\Sigma: \
  \big\{\xi \mapsto -\dd n(\xi)\big\} 
  \,.
$$
Denoting (a bit confusingly) by $x^1,\dots,x^{d-1}$
a local coordinate system of~$\Sigma$, 
$L$~can be identified with a $(1,1)$ mixed tensor 
having the matrix representation~$L^{\mu}_{\ \nu}$
with respect to the coordinate basis 
$(\partial_{x^1}, \dots, \partial_{x^{d-1}})$.
Here and in the sequel, we abbreviate 
$\partial_{x^\mu} := \partial/\partial x^\mu$
and assume the range of Greek indices being $1, \dots, d-1$.  
The relationship of~$L$ with 
the \emph{second fundamental form}~$h$ of~$\Sigma$
is through the formula
$
  L^{\mu}_{\ \nu} = g^{\mu\rho} h_{\rho\nu}
$,
where, as usual, $g^{\mu\nu}$ denote 
the entries of the inverse matrix $(g_{\mu\nu})^{-1}$
and the Einstein summation convention is employed.
For more details on the geometry of hypersurfaces,
we refer for instance to~\cite[Chap.~1]{Spivak2}.

The eigenvalues of~$L$ are called
\emph{principal curvatures} $\kappa_1 \dots \kappa_{d-1}$ of~$\Sigma$. 
It will be convenient to introduce the quantity
$$
  \varrho_{m}
  :=\big(\max\big\{
  \|\kappa_{1}\|_{\infty},\ldots,\|\kappa_{d-1}\|_{\infty}\big\}
  \big)^{-1}
  \,,
$$
with the convention that $\varrho_{m}=\infty$ 
if all $\kappa_{\mu}$ are identically equal to zero, 
and $\varrho_{m}=0$ if one of them is unbounded.
With help of the principal curvatures, 
we can construct $d-1$ invariants of~$L$:
$$
  K_{\mu}:=\binom{d-1}{\mu}^{-1}
  \sum_{\alpha_{1}<\ldots< \alpha_{\mu}}
  \kappa_{\alpha_{1}}\ldots\kappa_{\alpha_{\mu}}
$$
called $\mu$-th \emph{mean curvatures}~\cite{Kuhnel}.

Given $I:=(-1,1)$ and $\varepsilon>0$,
we define a \emph{layer}~$\Omega_\eps$ of width~$2\varepsilon$
along~$\Sigma$ as the image of the mapping
\begin{equation}\label{layer}
  \mathscr{L}:\Sigma\times I\to\R^d: \ 
  \big\{(x,u)\mapsto x+\varepsilon u n\big\}
  \,,
\end{equation}
\ie, $\Omega_\eps:=\mathscr{L}(\Sigma\times I)$.
We always assume that~$\Omega_\eps$ does not overlap itself, 
\ie, more precisely,
\begin{equation}\label{Ass.basic}
  \fbox{$
  \varepsilon<\varrho_{m}
  \qquad\text{and}\qquad	
  \mathscr{L} \quad \mbox{is injective}
  $.}
\end{equation}
Since~$\varepsilon$ is a small parameter in our setting,
the former will be always satisfied provided that the principal
curvatures are bounded, while the latter contains a non-trivial
hypothesis about the global geometry of~$\Sigma$.  

The relevance of our basic hypothesis~\eqref{Ass.basic} can be seen as follows.
The metric~$G$ induced by~\eqref{layer} has a block form
$$
  G = g \circ (I\!d-\varepsilon uL)^2 
  + \varepsilon^2 \, \dd u^2
  \,,
$$
where~$I\!d$ denotes the identity map on~$T_x\Sigma$.
In particular,
\begin{align*}
  |G|:=\det(G)
  &= \varepsilon^2 |g|\,\big[\det{(1-\varepsilon u L)}\big]^2
  = \varepsilon^2 |g|\,\prod_{\mu=1}^{d-1}(1-\varepsilon u\kappa_{\mu})^2
  \\
  & = \varepsilon^2 |g|\, \left[
  1+\sum_{\mu=1}^{d-1}(-\varepsilon u)^{\mu}\binom{d-1}{\mu}K_{\mu}
  \right]^2
  \,,
\end{align*}
with $|g|:=\det(g)$.
The formula implies that~$G$ is non-singular under 
the first condition in~\eqref{Ass.basic}.
Consequently, by the inverse function theorem, 
$\mathscr{L}:\Sigma\times I\to\Omega_\eps$ is a local diffeomorphism,
excluding thus ``local self-intersections''.
It will turn into a global diffeomorphism
under the additional injectivity hypothesis.
It then follows that, under the hypothesis~\eqref{Ass.basic},
$\Omega_\eps$~has indeed the geometrical meaning of the set of points
in~$\R^d$ squeezed between two parallel hypersurfaces 
at the distance~$\varepsilon$ from~$\Sigma$, \cf~\eqref{layer.intro}.
Furthermore, $\Omega_\eps$ can be identified 
with the Riemannian manifold $(\Sigma\times I,G)$.

\begin{remark}[On the injectivity assumption]
It is possible to consider $(\Sigma\times I,G)$ as an abstract
Riemannian manifold where only the hypersurface~$\Sigma$
is embedded (or even just immersed) in~$\R^{d}$. 
Then we do not need to assume the second part 
of hypothesis~\eqref{Ass.basic}. 
\end{remark}

Writing $x^d := u$ and 
$\partial_{x^d} := \partial/\partial u$ with $u \in I$,
$(\partial_{x^1},\dots,\partial_{x^d})$ represents a natural
coordinate frame for~$\Omega_\eps$.
We thus extend our index convention by assuming 
the range of Latin indices being $1,\dots,d$.
Then the metric~$G$ can be written 
in the following matrix representation
\begin{equation}
  (G_{ij})=
  \begin{pmatrix}
         (G_{\mu\nu}) & 0\\0 & \varepsilon^2
  \end{pmatrix},
  \qquad 
  G_{\mu\nu}=g_{\mu\rho}
  (\delta^{\rho}_{\sigma} -\varepsilon u L^{\rho}_{\ \sigma})
  (\delta^{\sigma}_{\nu} -\varepsilon u L^{\sigma}_{\ \nu})
  \,.
\end{equation}
We also introduce $(G^{ij}) := (G_{ij})^{-1}$
and the volume element 
$$
  \dd\Omega_\eps := |G|^{1/2} \, \dd\Sigma \wedge \dd u
  \,,
$$
where $\dd\Sigma := |g|^{1/2} \, \dd x^1 \wedge \dots \wedge \dd x^{d-1}$.
The reader is warned that,
in order not to additionally burden the index notation,
we do not make the dependence of~$G$ on~$\varepsilon$ 
explicit in this letter,
and similarly for several other quantities appearing in the text.
We have
\begin{equation}\label{eq:metric_bound}
  C_{-}(g_{\mu\nu})\leq(G_{\mu\nu})\leq C_{+}(g_{\mu\nu})
  \qquad \mbox{with} \qquad
  C_{\pm}:=(1\pm\varepsilon\varrho_{m}^{-1})^2=1+\mathcal{O}(\varepsilon)  
\end{equation}
as $\varepsilon\to 0$.

\section{Magnetic field in curvilinear coordinates}\label{Sec.field}
%
The identification of~$\Omega_\eps$ with~$(\Sigma\times I,G)$
can be understood as expressing the former in suitable (local)
curvilinear coordinates, namely $x \equiv (x^1, \dots, x^{d})$. 
The aim of this section is to introduce a covariant framework
for dealing with the magnetic field in these coordinates.

In any dimension, we introduce the ``magnetic field'' 
through its vector potential.
Let 
$
  A \equiv (A_1,\dots,A_d):
  \R^d \to \R^d
$
be a $C^1$-smooth function and let us call it a \emph{vector potential}
expressed in Cartesian coordinates $y\equiv(y^1, \dots, y^d)$ of~$\R^d$.
It gives rise to a 1-form
\begin{equation}\label{eq:vec_pot_form}
  \alpha = A_{i} \, \dd y^i
  \,.
\end{equation}
Passing locally to other coordinates 
\begin{equation}\label{eq:coord_trafo}
  y=\Phi(x)
  \,,
\end{equation}
then using the pull-back, the form transforms as follows
$$
  \Phi^{*}\alpha=\tilde{A}_{i} \, \dd x^i
  \qquad\text{with}\qquad 
  \tilde{A} := (D\Phi)^{T} A\circ\Phi
  \,,
$$
just because
$
  \dd y^{i}=\frac{\partial y^i}{\partial x^j} \, \dd x^j
$.
This is the way how we express~$A$ 
in local coordinates of~$(\Sigma\times I,G)$,
using particular charts of~$\Sigma$. 

The \emph{magnetic field} is then introduced as
the antisymmetric 2-form 
\begin{equation}\label{exact}
  \beta := \dd \alpha
  \,.
\end{equation}
Remark that we have some freedom in the choice of $\alpha$
to get the same~$\beta$,
which is the well known choice of gauge.
Indeed, if $\phi$ stands for a differentiable scalar function, 
then $\dd(\alpha+\dd\phi)=\dd\alpha=\beta$.  
In the Cartesian coordinates~$y$, we have
$
  \beta_{ij}=\partial_{y^i}A_{j}-\partial_{y^j}A_{i}
$,
and similarly for any (curvilinear) coordinates~$x$.

\subsection{Physical realizations}\label{Sec.subfield}
Now we explain how the general framework is related 
to the physical notion of magnetic field
in low-dimensional Euclidean spaces.

\subsubsection{Case of $d=3$}
The physical object is the magnetic field (magnetic induction) which we identify with the 1-form
$
  B = B_{i} \, \dd y^i = \tilde{B}_{i} \, \dd x^i
$. It obeys the second Maxwell equation (Gauss' law for magnetism) that reads
\begin{equation}\label{Gauss-law}
  *\dd*B=0 
  \,,
  \qquad\ie\quad \dd*B=0
  \,,
\end{equation}
where~$*$ stands for the Hodge star operator.
Recall that on a three-dimensional
Riemannian manifold with a metric tensor $G$, 
we have
$$
  *\dd x^i=\frac{1}{2}\, |G|^{1/2} \, G^{il} \, \delta^{123}_{lkm} \,
\dd x^k \wedge\dd x^m
\,, \quad
*(\dd x^i \wedge\dd x^j)=|G|^{1/2} \, G^{il} \, G^{jk} \, 
\delta^{123}_{lkm} \, \dd x^m
\,,
$$
where $\delta^{123}_{lkm}$ is 
the generalized Kronecker symbol~\cite{nakahara}.
For manifolds equipped with the flat metric, 
$\dd$, $*\dd *$, and $*\dd$ correspond 
to the usual $\grad$, $\divergence$, and $\curl$ operators, 
respectively.

We construct from~$B$ a 2-form $\beta:=*B$.
It follows from~\eqref{Gauss-law} that~$\beta$ is closed, 
\ie~$\dd\beta=0$. 
Since the ambient space is a Euclidean space,
it follows from the Poincar\'{e} lemma \cite[Chap.~7]{Spivak1}
that~$\beta$ is in fact exact. 
That is, \eqref{exact}~holds with some $1$-form~$\alpha$. 
Given a 2-form~\eqref{exact}, the magnetic field~$B$
can be reconstructed by the formula $B=*\beta=*\dd \alpha=\curl A$,
employing the fact that $**$~is an identity
in the present case.

In the Cartesian coordinates, we have
\begin{equation}\label{beta-form}
  \beta = \beta_{23}\,\dd y^2 \wedge\dd y^3 
  +\beta_{31}\,\dd y^3 \wedge\dd y^1
  +\beta_{12}\,\dd y^1 \wedge\dd y^2
  \,.
\end{equation}
The triple (not a vector!)
$$
  \gamma 
  :=(\beta_{23},\beta_{31},\beta_{12})
$$
transforms under the change of coordinates~(\ref{eq:coord_trafo})
as follows
\begin{equation}\label{eq:induction_trafo}
  \tilde{\gamma} := (\tilde{\beta}_{23},\tilde{\beta}_{31},\tilde{\beta}_{12})
  =\det{(D\Phi)}\,(D\Phi)^{-1}\,\gamma
  \,.
\end{equation}

\subsubsection{Case of $d=2$}
In the planar case, we have
$
  \dd\alpha=\tilde{\beta}_{12} \, \dd x^1 \wedge \dd x^2
$ 
and
\begin{equation}\label{eq:2d_mag_field}
  \tilde{B}
  :=*\dd\alpha=\tilde{\beta}_{12} \, 
  |G|^{1/2}\,G^{1j}\,G^{2k}\,\delta^{12}_{jk}
  =\tilde{\beta}_{12} \, |G|^{-1/2},
\end{equation}
gives the magnitude of the magnetic field in the direction
perpendicular to the plane. 

\section{The magnetic Laplacian}\label{Sec.Laplace}
%
Assume $A\in C^1(\R^d;\R^d)$.
Recall that~$\Omega_\eps$ is an open subset of~$\R^d$
as a consequence of~\eqref{Ass.basic}.
The quadratic form 
\begin{equation}\label{eq:sob_norm}
  Q[\psi] :=
  \big\|(-i\nabla+A)\psi\big\|_{L^2(\Omega_\eps)}^2 
\end{equation}
initially defined on $C_{0}^{\infty}(\Omega_\eps)$ is closable;
let us denote by the same symbol~$Q$ its closure. 
For bounded~$A$, it is easy to see that $\dom{Q}=H_0^1(\Omega_\eps)$. 
In general, however, we only have 
$
  \psi \in \dom{Q}
  \Rightarrow 
  |\psi| \in H_0^1(\Omega_\eps)
$,
which is a consequence of the diamagnetic inequality~\cite[Thm.~7.21]{LL}.
We define the \emph{magnetic Laplacian} $-\Delta_{D,A}^{\Omega_\eps}$
as the unique self-adjoint operator associated with the closure~$Q$
via the first representation theorem \cite[Thm.~VI.2.6]{Kato}.

%
\begin{remark}[On the regularity of~$A$]
For the definition of $-\Delta_{D,A}^{\Omega_\eps}$
via the quadratic form~\eqref{eq:sob_norm},
it is enough to assume $A\in L_\mathrm{loc}^2(\R^d;\R^d)$.
Our smoothness hypothesis will be needed later, \cf~\eqref{Taylor}.
\end{remark}

As explained in Section~\ref{Sec.field},
let us denote by~$\tilde{A}$ the components of the vector potential
expressed in the curvilinear coordinates induced by the embedding~\eqref{layer}.
Moreover, assume 
\begin{equation}\label{gauge}
  \tilde{A}_{d} = 0 
  \,.
\end{equation}
This may be always achieved by using an appropriate gauge transform, namely,
$$
  \tilde{A}
  \,\mapsto\,
  \tilde{A}-\nabla\int_{0}^{u}\tilde{A}_{d}(x^\mu,t) \, \dd t
  \,.
$$
Note that after this gauge transform, $\tilde{A}$ is continuous and it
has a continuous derivative in the variable $u$.

Employing the diffeomorphism $\mathscr{L}:\Sigma \times I \to \Omega_\eps$, 
we may thus identify $-\Delta_{D,A}^{\Omega_\eps}$ 
with an operator $\hat{H}$ on $L^{2}(\Sigma\times I,\dd\Omega_\eps)$ 
that acts, in the form sense, as 
\begin{equation*}
  \hat{H}
  =|G|^{-1/2}(-i\partial_{x^\mu}+\tilde{A}_{\mu})
  |G|^{1/2}G^{\mu\nu}(-i\partial_{x^\nu}
  +\tilde{A}_{\nu})-\varepsilon^{-2}|G|^{-1/2}
  \partial_{u}|G|^{1/2}\partial_{u}
  \,.
\end{equation*}
More precisely, $\hat{H}$~is the operator associated 
on $L^{2}(\Sigma\times I,\dd\Omega_\eps)$ with the quadratic form
$\hat{h}[\psi] := Q[\hat{U}^{-1}\psi]$, 
$\dom{\hat{h}} := \hat{U} [\dom{Q}]$,
where $\hat{U}\psi := \psi \circ \mathscr{L}$.

For our purposes, it will be more convenient to work 
with a unitarily equivalent operator on a Hilbert space
\emph{independent of~$\varepsilon$}.
Let us define 
$$
  J := \frac{1}{4} \ln\frac{|G|}{|g|}
  = \frac{1}{2}\sum_{\mu=1}^{d-1}\ln(1-\varepsilon u\kappa_{\mu})
  =\frac{1}{2}\ln\left[1+\sum_{\mu=1}^{d-1}(-\varepsilon u)^{\mu}
  \binom{d-1}{\mu}K_{\mu}\right]
  .
$$
Using the unitary transform 
$$
  U:L^{2}(\Sigma\times I,\dd\Omega_\eps)
  \to L^{2}(\Sigma\times I,\dd\Sigma \wedge \dd u) : \
  \left\{
  \psi\mapsto \ee{J} \psi
  \right\}
  \,,
$$ 
we arrive at the unitarily equivalent operator 
\begin{equation*}
  H: = U\hat{H}U^{-1}
  = |g|^{-1/2}(-i\partial_{x^\mu}+\tilde{A}_{\mu})
  |g|^{1/2}G^{\mu\nu}(-i\partial_{x^\nu}+\tilde{A}_{\nu})
  -\varepsilon^{-2}\partial_{u}^{2}+V
  \,,
\end{equation*}
where
$$
  V := |g|^{-1/2} \, 
  \partial_{x^i}\big( |g|^{1/2} G^{ij} (\partial_{x^j} J) \big)
  + (\partial_{x^i} J) G^{ij} (\partial_{x^j} J)
  \,.
$$
Again, the expressions should be understood in the sense of forms.
$H$~is the operator associated 
on $L^{2}(\Sigma\times I,\dd\Sigma \wedge \dd u)$ 
with the quadratic form
$h[\psi] := \hat{h}[U^{-1}\psi]$, 
$\dom{h} := U [\dom{\hat{h}}]$.
Summing up, 
$$
  H = U\hat{U}(-\Delta_{D,A}^{\Omega_\eps})\hat{U}^{-1}U^{-1}
  \,.
$$

\begin{remark}[On the regularity of~$\Sigma$]
Note that while~$\hat{H}$ can be introduced under 
the conventional $C^2$~smoothness assumption about~$\Sigma$,
it is because of the operator~$H$ why we actually require~$C^3$.  
Indeed, we need to differentiate the principal curvatures of~$\Sigma$
appearing in~$J$ in order to define~$H$, 
even if this is understood as an operator associated 
with the quadratic form~$h$.  
To proceed without the additional regularity,
one can apply the recent idea of refined 
$\eps$-dependent smoothing of curvatures~\cite{KSed},
but the overall analysis would become much more cumbersome.
\end{remark}

Henceforth, we work in the $\varepsilon$-independent
Hilbert space $L^{2}(\Sigma\times I,\dd\Sigma \wedge \dd u)$,
the norm and inner product of which will be denoted
by $\|\cdot\|$ and $\langle\cdot,\cdot\rangle$, respectively.
The norm and inner product in $L^{2}(\Sigma,\dd\Sigma)$
will be denoted
by $\|\cdot\|_g$ and $\langle\cdot,\cdot\rangle_g$, respectively.

\section{Comparison operators}\label{Sec.compar}
%
Applying~(\ref{eq:metric_bound}) to 
$
  \big\langle 
  (\partial_{x^\mu} J)\psi-(\partial_{x^\mu}+iA_{\mu})\psi,
  G^{\mu\nu}[(\partial_{x^\nu} J)\psi
  -(\partial_{x^\nu}+iA_{\nu})\psi]
  \big\rangle
  \,,
$ 
with $\psi \in C_0^1(\Sigma \times I)$,
we obtain crucial bounds
\begin{equation}\label{bound1}
  H_{-}\leq H\leq H_{+}
  \,,
\end{equation}
in the form sense, with the comparison operators
\begin{equation*}
   H_{\pm}
  :=C^{-1}_{\mp}\left(|g|^{-1/2}(-i\partial_{x^\mu}+\tilde{A}_{\mu})
  |g|^{1/2}g^{\mu\nu}(-i\partial_{x^\nu}+\tilde{A}_{\nu})
  +v_{1}\right)-\varepsilon^{-2}\partial_{u}^{2}+V_{2}
  \,,
\end{equation*}
where
\begin{align*}
 v_{1}&:=\Delta_{g}J+|\nabla_{\!g}J|_{g}^{2}
  = \frac{1}{2}\frac{\Delta_{g}f}{1+f}
  -\frac{1}{4}\frac{|\nabla_{\!g}f|^{2}_{g}}{(1+f)^2}
  \,, \qquad 
  f := \sum_{\mu=1}^{d-1}(-\varepsilon u)^{\mu}\binom{d-1}{\mu}K_{\mu}
  \,,
  \\
 V_{2}&:=\varepsilon^{-2}
  \left[ \partial_u^2 J + (\partial_u J)^2 \right]
  =-\frac{1}{2}\sum_{\mu=1}^{d-1}
  \frac{\kappa_{\mu}^{2}}{(1-\varepsilon u\kappa_{\mu})^2}
  +\frac{1}{4}\left(\sum_{\mu=1}^{d-1}
  \frac{\kappa_{\mu}}{1-\varepsilon u\kappa_{\mu}}\right)^2
  \,.
\end{align*}
Here we have introduced coordinate-free notations
$$
  \Delta_{g}f
  :=|g|^{-1/2}\partial_{x^\mu}(|g|^{1/2}g^{\mu\nu}\partial_{x^\nu}f)
  \,, \qquad
  |\nabla_{\!g}f|_{g} := 
  \sqrt{ (\partial_{x^\mu}f) g^{\mu\nu} (\partial_{x^\nu}f) }
  \,.
$$
In order to give a meaning to the Laplacian of curvatures
in the definition of the potential~$v_1$,
we need to strengthen our regularity assumptions about~$\Sigma$. 
Henceforth, we assume
\begin{equation}\label{Ass.C4}
  \fbox{$
  |\nabla_{\!g}\kappa_{\mu}|_{g}
  \,, \
  \Delta_{g}\kappa_{\mu}
  \in L^{\infty}(\Sigma)
  $,}
\end{equation}
which is equivalent to
$
  |\nabla_{\!g}K_{\mu}|_{g},
  \Delta_{g}K_{\mu}, 
  \in L^{\infty}(\Sigma)
$.
Recall that the assumption 
$\kappa_{\mu} \in L^{\infty}(\Sigma)$
is implicit in the first condition of~\eqref{Ass.basic}.
Under the hypotheses, we have
$$ 
  v_{1}=\mathcal{O}(\varepsilon)
  \,, \qquad 
  V_{2}=V_{\mathrm{eff}}
  +\mathcal{O}(\varepsilon)
$$
uniformly as $\varepsilon\to 0$,
where
\begin{equation}\label{V.eff}
  V_{\mathrm{eff}}
  :=-\frac{1}{2}\sum_{\mu=1}^{d-1}\kappa_{\mu}^{2}
  +\frac{1}{4}\left(\sum_{\mu=1}^{d-1}\kappa_{\mu}\right)^2
  \,.
\end{equation}

From~\eqref{bound1} we deduce cruder bounds
\begin{equation}\label{eq:form_bound}
  H^{-}\leq H\leq H^{+}
\end{equation}
with
\begin{equation*}
  H^{\pm}
  :=C^{-1}_{\mp}
  \left(
  |g|^{-1/2}(-i\partial_{x^\mu}+\tilde{A}_{\mu})
  |g|^{1/2}g^{\mu\nu}(-i\partial_{x^\nu}+\tilde{A}_{\nu})
  +V_{\mathrm{eff}}
  \right)
  -\varepsilon^{-2}\partial_{u}^{2}
  \pm C_{0}
  \,,
\end{equation*}
where $C_{0}$ is a positive constant 
such that $C_{0}=\mathcal{O}(\varepsilon)$
as $\varepsilon\to 0$.
Remark that this result was obtained earlier 
in \cite{DEK2,EK3} for the special case $d=3$ and $A=0$.

If $\tilde{A}_{\mu}$ is independent 
of the ``transverse'' coordinate~$u$, 
then $H^{\pm}$ are decoupled on
$L^{2}(\Sigma,\dd\Sigma)\otimes L^{2}(I,\dd u)$. 
Our next aim will be to find decoupled comparison operators 
in the general case.
To this purpose define an $\varepsilon$-independent quantity 
$\hat{A}(x,t):=\tilde{A}(x,t/\varepsilon)$ 
and expand it in the last variable into the Taylor series,
\begin{equation}\label{Taylor}
  \tilde{A}(x,u)=\hat{A}(x,\varepsilon u)
  =\hat{A}(x,0)+\varepsilon u \, \partial_{d}\hat{A}(x,\xi(x,u))
  =:\tilde{A}(x,0)+\varepsilon A'(x,u)
  \,,
\end{equation}
with some $\xi(x,u)\in(-\varepsilon,\varepsilon)$. 
It is important to stress that~$A'$ depends on~$\eps$
only through~$\xi$, which measures  
the actual distance from~$\Sigma$ in~$\R^d$. 

For any trial function $\psi \in C_0^1(\Sigma \times I)$, 
we have
\begin{eqnarray*}
\lefteqn{
 \left\langle(-i\partial_{x^\mu}+\tilde{A}_{\mu})\psi,g^{\mu\nu}
  (-i\partial_{x^\nu}+\tilde{A}_{\nu})\psi\right\rangle_{\!g}
}
  \\
  &&=\left\langle
  \big(-i\partial_{x^\mu}+\tilde{A}_{\mu}(\cdot,0)\big)\psi,
  g^{\mu\nu}\big(-i\partial_{x^\nu}+\tilde{A}_{\nu}(\cdot,0)\big)\psi
  \right\rangle_{\!g} 
  \\  
  &&\phantom{=} +2\varepsilon\Re
  \left\langle A'_{\mu}\psi,
  g^{\mu\nu}\big(-i\partial_{x^\nu}+\tilde{A}_{\nu}(\cdot,0)\big)\psi
  \right\rangle_{\!g}
  +\varepsilon^2
  \big\langle A'_{\mu}\psi,
  g^{\mu\nu}A'_{\nu}\psi
  \big\rangle_{g}
  \,.
\end{eqnarray*}
By the Cauchy-Schwarz and the Young inequalities,
\begin{multline*}
  \left|\left\langle 
  A'_{\mu}\psi,g^{\mu\nu}
  (-i\partial_{x^\nu}+\tilde{A}_{\nu}(.,0))\psi
  \right\rangle_{\!g}
  \right|
  \\ 
  \leq \frac{1}{2}
  \big\langle 
  A'_{\mu}\psi,g^{\mu\nu}A'_{\nu}\psi
  \big\rangle_{g}
  +\frac{1}{2}
  \left\langle\big(-i\partial_{x^\mu}+\tilde{A}_{\mu}(.,0)\big)\psi,
  g^{\mu\nu}\big(-i\partial_{x^\nu}+\tilde{A}_{\nu}(.,0)\big)\psi
  \right\rangle_{\!g}
  \,.
\end{multline*}
Consequently, 
$$
  h^{-} \leq 
  |g|^{-1/2}(-i\partial_{x^\mu}+\tilde{A}_{\mu})
  |g|^{1/2}g^{\mu\nu}
  (-i\partial_{x^\nu}+\tilde{A}_{\nu})
  \leq h^{+}
  \,,
$$
where
$$
  h^{\pm}
  :=(1\pm\varepsilon)|g|^{-1/2}
  \big(-i\partial_{x^\mu}+\tilde{A}_{\mu}(.,0)\big)
  |g|^{1/2}g^{\mu\nu}
  \big(-i\partial_{x^\nu}+\tilde{A}_{\nu}(.,0)\big)
  +(\pm\varepsilon+\varepsilon^2)A'_{\mu}g^{\mu\nu}A'_{\nu}
  \,.
$$
Putting this into~(\ref{eq:form_bound}), 
we arrive at the following result.
\begin{proposition}\label{prop:weak_convergence}
In addition to~\eqref{Ass.basic} and~\eqref{Ass.C4},
let us assume 
\begin{equation}\label{Ass.A}
  \fbox{$
  A'_{\mu}\,g^{\mu\nu}A'_{\nu}
  \in L^{\infty}\big(\Sigma\times (-\varrho_m,\varrho_m)\big)
  $,}
\end{equation}
where~$A'_{\mu}$ is introduced in~\eqref{Taylor}.
Then
\begin{equation}\label{eq:final_comparison}
  H^{-}_{0}\leq H\leq H^{+}_{0}
\end{equation}
with
\begin{align*}
  H^{\pm}_{0}:=\ &\mathcal{C}_{\pm}
  \left(|g|^{-1/2}\big(-i\partial_{x^\mu}+\tilde{A}_{\mu}(.,0)\big)
  |g|^{1/2}g^{\mu\nu}\big(-i\partial_{x^\nu}+\tilde{A}_{\nu}(.,0)\big)
  +V_{\mathrm{eff}}\right)
  \\
  \ &-\varepsilon^{-2}\partial_{u}^{2}\pm \mathcal{C}_{0}
  \,,
\end{align*}
where 
$
  \mathcal{C}_{\pm}
  :=(1\pm\varepsilon)C_{\mp}^{-1}
  =1+\mathcal{O}(\varepsilon)
$ 
and $\mathcal{C}_{0}=\mathcal{O}(\varepsilon)$
as $\varepsilon \to 0$.
\end{proposition}
\begin{remark}
If $d=3$, then $\tilde{B}_{\mu}\in L^{\infty}(\Sigma\times I)$ 
implies 
$A'_{\mu}\in L^{\infty}(\Sigma\times (-\varrho_m,\varrho_m))$.
\end{remark}

Proposition~\ref{prop:weak_convergence}
suggests that for small values of $\varepsilon$, 
$H$~behaves like
\begin{equation}\label{H0}
  H_{0}=h_{\mathrm{eff}}-\varepsilon^{-2}\partial_{u}^{2}
  \ \simeq \ h_{\mathrm{eff}} \otimes 1 
  + 1 \otimes (-\varepsilon^{-2}\partial_{u}^{2})
\end{equation}
on 
$
  L^{2}(\Sigma \times I,\dd\Sigma\wedge du)
  \simeq
  L^{2}(\Sigma,\dd\Sigma)\otimes L^{2}(I,\dd u)
$
with the \emph{effective Hamiltonian}
\begin{equation}\label{H.eff}
  h_{\mathrm{eff}}
  := |g|^{-1/2}\big(-i\partial_{x^\mu}+\tilde{A}_{\mu}(.,0)\big)
  |g|^{1/2}g^{\mu\nu}\big(-i\partial_{x^\nu}+\tilde{A}_{\nu}(.,0)\big)
  +V_{\mathrm{eff}}
  \,.
\end{equation}
In particular, 
applying the minimax principle to~(\ref{eq:final_comparison}), 
one can show that any eigenvalue of~$H$ 
that lies below the essential spectrum 
is well approximated by an eigenvalue of~$H_{0}$ 
for~$\varepsilon$ small enough. 
To say more about the convergence of the spectrum 
and the eigenfunctions, 
we need the norm resolvent convergence of the respective operators,
which we establish in the next section.

\section{The norm-resolvent convergence}\label{Sec.nrs}
%
As $\varepsilon\to 0$, the second term of~$H_{0}$ diverges
and consequently the spectrum of~$H_{0}^{\pm}$ explodes.
To renormalize it, let us introduce~$E_{m}$ 
as the $m$-th eigenvalue of $-\partial_{u}^{2}$ on $L^2(I)$,
subject to Dirichlet boundary condition, 
\ie~$E_{m}=(m\pi/2)^2$ with $m\in\N$, 
and subtract $\varepsilon^{-2}E_{1}$ 
from both~$H_{0}^{\pm}$ and~$H$:
$$
  H_{\mathrm{ren}}:=H-\varepsilon^{-2}E_{1}
  \,, \qquad 
  H_{0,\mathrm{ren}}:=H_{0}-\varepsilon^{-2}E_{1}
  \,.
$$
With this renormalization, 
\eqref{eq:final_comparison}~reads
\begin{equation}\label{bound.ren}
  \mathcal{C}_{-}
  h_{\mathrm{eff}}-\varepsilon^{-2}(\partial_{u}^{2}+E_{1})
  -\mathcal{C}_{0}
  \leq H_{\mathrm{ren}} \leq 
  \mathcal{C}_{+}
  h_{\mathrm{eff}}-\varepsilon^{-2}(\partial_{u}^{2}
  +E_{1})+\mathcal{C}_{0}
  \,.
\end{equation}
Next, choose a constant~$k$ large enough 
that for all~$\varepsilon$ smaller than some $\varepsilon_{0}>0$, 
$$
  h_{\mathrm{eff}}+k \geq 1
  \qquad\mbox{and}\qquad
  \mathcal{C}_{-}h_{\mathrm{eff}}-\mathcal{C}_{0}+k \geq 1
  \,.
$$
Such~$k$ always exists, since $V_{\mathrm{eff}}$ is bounded.
Consequently, 
$
  \|(H_{\mathrm{ren}}+k)^{-1}\| \leq 1
$
and
$
  \|(H_{0,\mathrm{ren}}+k)^{-1}\| \leq 1
$.

Using \cite[Thm.~VI.2.21]{Kato}, 
we deduce from~\eqref{bound.ren}
the resolvent bounds
\begin{multline}\label{eq:fund_form_bound}
  [\mathcal{C}_{+}h_{\mathrm{eff}}
  -\varepsilon^{-2}(\partial_{u}^{2}+E_{1})
  +\mathcal{C}_{0}+k]^{-1}-(H_{0,\mathrm{ren}}+k)^{-1}
  \\
  \leq\ (H_{\mathrm{ren}}+k)^{-1}-(H_{0,\mathrm{ren}}+k)^{-1} \ \leq 
  \\
  [\mathcal{C}_{-}h_{\mathrm{eff}}
  -\varepsilon^{-2}(\partial_{u}^{2}+E_{1})
  -\mathcal{C}_{0}+k]^{-1}-(H_{0,\mathrm{ren}}+k)^{-1}
  \,.
\end{multline}

Our next strategy is to apply the following pair of observations:
\begin{lemma}\label{lem:form_perturbation}
Let~$T$ be a positive self-adjoint operator 
and let~$S$ stands for a symmetric operator 
that is relatively form bounded by~$T$ 
with the relative bound $a<1$ 
and the other constant $b=0$. 
Then $T^{-1/2}ST^{-1/2}$ (as a quadratic form) 
corresponds to a bounded operator~$L$. 
Moreover $\|L\|<1$ and
$$
  (T\dotplus S)^{-1}=T^{-1/2}(1+L)^{-1}T^{-1/2}
  \,.
$$
This immediately implies
\begin{equation}\label{eq:res_estimate}
  \|(T\dotplus S)^{-1}-T^{-1}\|\leq\frac{\|T^{-1}\|\|L\|}{1-\|L\|}
  \,.
\end{equation}
\end{lemma}
\begin{lemma}\label{lem:sandwich}
Let $\{L_{n}\},\{L_{n}^{\pm}\}$ 
be sequences of bounded self-adjoint operators and
$$
  L_{n}^{-}\leq L_{n}\leq L_{n}^{+}
  \,.
$$
Then $\|L_{n}\|\leq\max\{\|L_{n}^{+}\|,\|L_{n}^{-}\|\}$. 
In particular, if $\lim_{n\to\infty}\|L_{n}^{\pm}\|=0$, 
then also $\lim_{n\to\infty}\|L_{n}\|=0$.
\end{lemma}
\noindent
Lemma~\ref{lem:form_perturbation}
is just a special case of \cite[Thm.~6.25]{Teschl}
(see also~\cite{SiQF} for similar manipulations),
while Lemma~\ref{lem:sandwich} is a direct consequence
of the minimax principle.

In Lemma~\ref{lem:form_perturbation} put 
$$
  T := H_{0,\mathrm{ren}}+k
  \,, \qquad 
  S := H_{0}^{-}-H_{0}=(\mathcal{C}_{-}-1)h_{\mathrm{eff}}-\mathcal{C}_{0}
$$
(here we view $h_\mathrm{eff}$ as $h_\mathrm{eff}\otimes 1$
on $L^{2}(\Sigma,\dd\Sigma)\otimes L^{2}(I,\dd u)$). 
Then for all $\psi\in \dom{T^{1/2}}$, we have
\begin{equation*}
 \begin{split}
  |\langle\psi,S\psi\rangle|
  &=|\langle\psi,[(\mathcal{C}_{-}-1)(h_{\mathrm{eff}}+k)
  +(1-\mathcal{C}_{-})k-\mathcal{C}_{0}]\psi\rangle|
  \\
  &\leq |\mathcal{C}_{-}-1|
  \langle\psi,(h_{\mathrm{eff}}+k)\psi\rangle
  +|(1-\mathcal{C}_{-})k-\mathcal{C}_{0}|\langle\psi,\psi\rangle
  \\
  &\leq \left( 1-\mathcal{C}_{-}+|(1-\mathcal{C}_{-})k
  -\mathcal{C}_{0}|\|T^{-1}\|\right)\langle\psi,T\psi\rangle
  \,.
 \end{split}
\end{equation*}
Here $\langle\psi,S\psi\rangle$ is understood 
as the action on $\psi$ of the quadratic form associated with~$S$ 
and similarly for the other operators. 
In the second inequality we have used that 
$h_{\mathrm{eff}}\leq H_{0,\mathrm{ren}}$.
Since
$$
  a_{-}:=1-\mathcal{C}_{-}+|(1-\mathcal{C}_{-})k
  -\mathcal{C}_{0}|\|(H_{0,\mathrm{ren}}+k)^{-1}\|
  =\mathcal{O}(\varepsilon)
  \quad\text{as}\quad\varepsilon\to 0
  \,,
$$
we see that $a_{-}<1$ for $\varepsilon$ small enough, 
and from~(\ref{eq:res_estimate}) we infer that
\begin{equation*}
  \left\|
  \left(\mathcal{C}_{-}h_{\mathrm{eff}}
  -\varepsilon^{-2}(\partial_{u}^{2}+E_{1})
  -\mathcal{C}_{0}+k\right)^{-1}-(H_{0,\mathrm{ren}}+k)^{-1}
  \right\|
  \leq\frac{\|(H_{0,\mathrm{ren}}+k)^{-1}\|~a_{-}}{1-a_{-}}
  \,.
\end{equation*}
In a similar manner we obtain that whenever
$$
  a_{+}:=\mathcal{C}_{+}-1+|(1-\mathcal{C}_{+})k
  +\mathcal{C}_{0}|\|(H_{0,\mathrm{ren}}+k)^{-1}\|
  =\mathcal{O}(\varepsilon)<1
  \,,
$$
then
\begin{equation*}
  \left\|
  \left(\mathcal{C}_{+}h_{\mathrm{eff}}
  -\varepsilon^{-2}(\partial_{u}^{2}+E_{1})
  +\mathcal{C}_{0}+k\right)^{-1}-(H_{0,\mathrm{ren}}+k)^{-1}
  \right\|
  \leq\frac{\|(H_{0,\mathrm{ren}}+k)^{-1}\|~a_{+}}{1-a_{+}}
  \,.
\end{equation*}
Putting these two estimates together 
with Lemma~\ref{lem:sandwich} and~(\ref{eq:fund_form_bound}), 
we arrive at the key result of this paper:
\begin{theorem}\label{Thm.main}
Assume~\eqref{Ass.basic}, \eqref{Ass.C4} and \eqref{Ass.A}.
Then
\begin{align*}
 \|(H_{\mathrm{ren}}+k)^{-1}-(H_{0,\mathrm{ren}}+k)^{-1}\|
  &\leq\|(H_{0,\mathrm{ren}}+k)^{-1}\|
  \max\left\{\frac{a_{-}}{1-a_{-}},\frac{a_{+}}{1-a_{+}}\right\}
  \\
  &=\mathcal{O}(\varepsilon)
  \quad\text{as}\quad\varepsilon\to 0
  \,.
\end{align*}
\end{theorem}
\begin{remark}[Gauge invariance]
Recall that we have worked in a special gauge,
namely~\eqref{gauge}. 
If we had started with another one that differs by $\nabla \phi$ 
for some real differentiable function~$\phi$, 
we would have obtain
$$
  -\Delta^{\Omega_\eps}_{D,A+\nabla \phi}
  =(-i\nabla+A+\nabla \phi)^2
  = \ee{-i\phi}(-\Delta^{\Omega_\eps}_{D,A}) \, \ee{i\phi}
  \,.
$$
This implies that when passing to this new gauge, 
$\hat{H}$ and~$H$ must be interchanged for 
$\ee{-i\tilde{\phi}}\hat{H}\ee{i\tilde{\phi}}$ and 
$\ee{-i\tilde{\phi}}H\ee{i\tilde{\phi}}$, respectively, 
where $\tilde{\phi}:=\phi \circ \mathscr{L}$. 
All the weak estimates for the comparison operators above 
may be sandwiched by $\ee{\mp i\tilde{\phi}}$ 
and they still remain valid. 
Consequently, we obtain the inequality of Theorem~\ref{Thm.main}
for the pair $\ee{-i\tilde{\phi}}H_{\mathrm{ren}}\ee{i\tilde{\phi}}$ and 
$\ee{-i\tilde{\phi}}H_{0,\mathrm{ren}}\ee{i\tilde{\phi}}$ 
with exactly the same upper bound. 
\end{remark}
\begin{remark}[Electric field]
If we had started with a full Schr\"odinger operator in~$\Omega_\eps$,
\ie, $-\Delta_{D,A}^{\Omega_\eps} + V$, 
where the scalar potential~$V$ that represents an ambient electric field
is such that $\partial_{y^j}V$ is bounded on $\Omega_{\varrho_{m}}$,  
we would have arrived at the same convergence result with 
the effective Hamiltonian~\eqref{H.eff} merely modified 
by adding the projection $(V\circ\mathscr{L})(\cdot,0)$.
This can be established quite straightforwardly with the aid of the following Taylor expansion
$$(V\circ\mathscr{L})(x,u)=(V\circ\mathscr{L})(x,0)+\varepsilon u n^j(x) (\partial_{y^j}V\circ\mathscr{L})(x,\xi)$$
that holds for any $x\in\Sigma$, and where $\xi=\xi(x,u)\in I$.

However, if $V$ is singular on $\Sigma$, the situation is much more delicate. For a model example see \cite{DuStTu}, where
a planar layer with the Coulomb potential and without any magnetic field was considered.
\end{remark}
%

\section{A dimensional reduction}\label{Sec.reduction}
%
In this section we derive a variant of Theorem~\ref{Thm.main}
by replacing $H_{0,\mathrm{ren}}$ directly by 
the $(d-1)$-dimensional effective Hamiltonian~$h_\mathrm{eff}$, 
\cf~\eqref{H.eff}.
It requires certain prerequisites and identifications,
because~$H_{\mathrm{ren}}$ and~$h_\mathrm{eff}$ 
act on different Hilbert spaces.

Let us denote by~$\chi_m$ the eigenfunction of~$-\partial_u^2$
(subject to Dirichlet boundary conditions)
corresponding to~$E_m$
In particular, we choose 
$$
  \chi_{1}(u) := \cos(\pi u/2)
  \,.
$$
We decompose our Hilbert space into an orthogonal sum
$$
  L^{2}(\Sigma\times I ,\dd\Sigma\wedge du)
  = \mathcal{H}_1 \oplus \mathcal{H}_1^\bot
  \,,
$$
where the subspace~$\mathcal{H}_1$ consists of functions of the form 
$$
  \psi_1(x^1,\dots,x^d)=\varphi(x^1,\dots,x^{d-1})\chi_1(x^d)
  \,.
$$
Given any $\psi \in L^{2}(\Sigma\times I ,\dd\Sigma\wedge du)$,
we have the decomposition with $\psi_1 \in \mathcal{H}_1$
and $\phi \in \mathcal{H}_1^\bot$.
More specifically, $\psi_1 = P_1\psi$ and $\phi = Q\psi$,
where 
$
  P_1:=\chi_{1} \, \langle\chi_{1},\cdot\rangle_{L^{2}(I)}
$
is the projection on the lowest transverse mode and $Q:=1-P_1$.
The mapping $\iota:\varphi\mapsto\psi_1$ is an isomorphism
of $L^2(\Sigma,d\Sigma)$ onto~$\mathcal{H}_1$.
Hence, with an abuse of notations, we may identify any operator~$h$
on $L^2(\Sigma,d\Sigma)$ with the operator $\iota h \iota^{-1}$
acting on $\mathcal{H}_1 \subset L^{2}(\Sigma\times I ,\dd\Sigma\wedge du)$.
Having this convention in mind, we can write 
$P_1H_{0,\mathrm{ren}}P_1 = h_{\mathrm{eff}}$ 
and have the following decomposition 
\begin{align*}
  H_{0,\mathrm{ren}}&=\begin{pmatrix}
         h_{\mathrm{eff}}&0\\0&QH_{0,\mathrm{ren}}Q
        \end{pmatrix}
  \,,
  \\
 (H_{0,\mathrm{ren}}+k)^{-1}&=\begin{pmatrix}
                 (h_{\mathrm{eff}}+k)^{-1}&0\\0&(QH_{0,\mathrm{ren}}Q+k)^{-1}
                \end{pmatrix}
  \,,
\end{align*}
as operators on $\mathcal{H}_1 \oplus \mathcal{H}_1^\bot$.

Since
$$
  Q(H_{0,\mathrm{ren}}+k)Q
  =Q(h_{\mathrm{eff}}+k)Q-\varepsilon^{-2}Q(\partial_{u}^{2}+E_{1})Q
  \geq\varepsilon^{-2}(E_{2}-E_{1})=\frac{3\pi^2}{4\varepsilon^2}
  \,,
$$
we obtain
\begin{equation*}
 \|(H_{0,\mathrm{ren}}+k)^{-1}-(h_{\mathrm{eff}}+k)^{-1}\oplus 0\|
  =\|(QH_{0,\mathrm{ren}}Q+k)^{-1}\|\leq\frac{4\epsilon^2}{3\pi^2}
  \,.
\end{equation*}
This together with Theorem~\ref{Thm.main} implies
\begin{theorem}\label{Thm.main.bis}
Under the hypotheses of Theorem~\ref{Thm.main},
\begin{equation*}
  \big\|(H_{\mathrm{ren}}+k)^{-1}-(h_{\mathrm{eff}}+k)^{-1}\oplus 0\big\|
  =\mathcal{O}(\varepsilon)
  \quad\text{as}\quad\varepsilon\to 0
  \,.
\end{equation*}
\end{theorem}

\subsection{Physical realizations}\label{Sec.phys}
The message of the general result is that 
the (unitarily transformed and suitably renormalized)
magnetic Laplacian~$-\Delta_{D,A}^{\Omega_\eps}$ in the layer~$\Omega_\eps$ 
behaves for small~$\varepsilon$ 
as the effective Hamiltonian~$h_\mathrm{eff}$
on the underlying hypersurface.
The coefficients of~$h_\mathrm{eff}$ depend on curvatures of~$\Sigma$
and on the ambient vector potential projected to~$\Sigma$,
\cf~\eqref{H.eff}.
Let us interpret this dependence
in physically interesting situations 
of low-dimensional Euclidean spaces.
We refer to Section~\ref{Sec.subfield}
for the notation concerning the magnetic field in our setting.

\subsubsection{Case of $d=3$}
In this case, 
$
  V_{\mathrm{eff}}=K_{2}-K_{1}^2
  =-\frac{1}{4}(\kappa_{1}-\kappa_{2})^{2}
$,
where $K_1=\frac{1}{2}(\kappa_1+\kappa_2)$ and $K_2=\kappa_1\kappa_2$
are the familiar mean and Gauss curvatures of~$\Sigma$, respectively.
Hence, if~$\Sigma$ is not a part of a plane or a sphere, 
then~$V_{\mathrm{eff}}$ always represents an \emph{attractive} interaction
(\ie~$V_{\mathrm{eff}}$ is non-positive and non-trivial).
The kinetic part of~$h_\mathrm{eff}$ is the magnetic
Laplace-Beltrami operator on the two-dimensional surface~$\Sigma$ 
with a vector potential associated with the 1-form
$$
  \alpha_{\mathrm{eff}} :=
  \tilde{A}_{1}(\cdot,0) \, \dd x^1 
  +\tilde{A}_{2}(\cdot,0) \, \dd x^2
  \,.
$$
Consequently, 
$
  \dd \alpha_{\mathrm{eff}} 
  = \tilde{\beta}_{12}(\cdot,0) \, \dd x^1 \wedge \dd x^2
$
with 
$
  \tilde{\beta}_{12}(\cdot,0) = 
  \partial_1 \tilde{A}_{2}(\cdot,0)
  - \partial_2 \tilde{A}_{2}(\cdot,0) 
$.
Using the transformation rule~(\ref{eq:induction_trafo}), 
we get
$$
  \tilde{\beta}_{12}|_{u=0}
  =\det(D\mathcal{L})|_{u=0}\,((D\mathcal{L})|_{u=0})^{-1}_{3j}\,B_{j}|_{u=0}
  =|g|^{1/2} \,n\cdot B|_{u=0}
  \,.
$$
Here the second equality is a consequence of a tedious computation
which can be greatly simplified by using the algebraic formula
\begin{equation*}
 M^{-1}=\frac{1}{\det{M}}
  \begin{pmatrix}
                     (\vec{m}_2\times \vec{m}_3)^{T}\\
		     (\vec{m}_3\times \vec{m}_1)^{T}\\
		     (\vec{m}_1\times \vec{m}_2)^{T}
                    \end{pmatrix}
  \qquad\mbox{for}\qquad 
  M:=(\vec{m}_1,\vec{m}_2,\vec{m}_3)
  \,, 
\end{equation*}
where $\vec{m}_i$ are three-dimensional column vectors. 
Finally, using~(\ref{eq:2d_mag_field}) for the two-dimensional metric~$g$, 
we conclude that the magnetic field associated with~$\alpha_{\mathrm{eff}}$
is given by
$$
  B_{\mathrm{eff}} := *\dd\alpha_{\mathrm{eff}} = n\cdot B|_{u=0}
  \,.
$$
Therefore in the limit $\varepsilon\to 0$, 
a particle confined to the layer~$\Omega_\eps$ 
is affected only by the projection of the original magnetic field 
into the direction that is normal to the underlying surface~$\Sigma$.

\subsubsection{Case of $d=2$}
In this case, 
$
  V_{\mathrm{eff}} =-\frac{1}{4} \kappa_{1}^{2}
$
is again attractive.
Unless~$\Sigma$ is a closed curve (\ie~homeomorphic to a circle),
there is no magnetic effect in the limit $\varepsilon \to 0$,
because the presence of $\tilde{A}_1(\cdot,0)$ 
in the one-dimensional kinetic part of~$h_\mathrm{eff}$
can be always gauged out.

%
\section{Spectral consequences}\label{Sec.spec}
%
We have obtained the norm-resolvent convergences of~$H_{\mathrm{ren}}$ 
to~$H_{0,\mathrm{ren}}$ or~$h_{\mathrm{eff}}$, respectively,
for a special value (namely~$-k$) of the spectral parameter 
from the resolvent set. 
However, for sufficiently small~$\varepsilon$ 
(the particular threshold value depends on the chosen value 
of the spectral parameter), these results may be extended
to the resolvent sets of~$H_{0,\mathrm{ren}}$ or~$h_{\mathrm{eff}}$, 
respectively, \cf~\cite[Eq.~IV.3.10]{Kato}. 

Now, let~$\mu$ be an isolated eigenvalue of~$h_{\mathrm{eff}}$
of finite multiplicity~$N$ and denote by 
$\varphi_1, \dots, \varphi_N$ the corresponding (orthogonal) eigenfunctions.
Let~$\Gamma$ be a contour in the complex plane  
of radius smaller than the isolation distance of~$\mu$
centered at~$\mu$. 
Then for all sufficiently small~$\varepsilon$, 
every point of~$\Gamma$ lies also 
in the resolvent set of~$H_{\mathrm{ren}}$.
Hence, it makes sense to define the eigenprojections
$$
  P
  :=-\frac{1}{2\pi i}\int_{\Gamma}(H_{\mathrm{ren}}-\xi)^{-1}\dd\xi
  \,, \qquad 
  P_{\mathrm{eff}}
  :=-\frac{1}{2\pi i}\int_{\Gamma}(h_{\mathrm{eff}}-\xi)^{-1}\dd\xi
  \,.
$$
Using Theorem~\ref{Thm.main.bis}, 
we obtain that $\|P-P_{\mathrm{eff}}\|=\mathcal{O}(\varepsilon)$ 
as $\varepsilon \to 0$.
In particular,  $\|P-P_{\mathrm{eff}}\| < 1$ for~$\varepsilon$ small enough. 
We conclude that for these values of $\varepsilon$, 
the number of eigenvalues of~$H_{\mathrm{ren}}$
(counting possible multiplicities)
lying inside~$\Gamma$ coincides with~$N$, \cf~\cite[Thm.~I.6.32]{Kato}.
Moreover, $\psi_{n}^\varepsilon:=P(\varphi_{n}\otimes\chi_1)$, 
with $n=1,\dots,N$,
are eigenfunctions of~$H_{\mathrm{ren}}$ associated with
these eigenvalues of~$H_{\mathrm{ren}}$,
which we denote by 
$
  \lambda_1^\varepsilon, \dots \lambda_N^\varepsilon
$
(they are not necessarily sorted in a non-decreasing order,
but they are counted according to multiplicities). 

\begin{corollary}\label{Thm.spec}
Under the hypotheses of Theorem~\ref{Thm.main}
and with the above notation, 
for every $n \in \{1,\dots,N\}$, we have
\begin{equation*}
  |\lambda_{n}^\varepsilon-\mu|=\mathcal{O}(\varepsilon)
  \qquad\mbox{and}\qquad
  \|\psi_{n}^\varepsilon-\varphi_{n}\otimes\chi_1\|=\mathcal{O}(\varepsilon)
  \qquad\text{as}\quad\varepsilon\to 0
  \,.
\end{equation*}
\end{corollary}

Note that the eigenvalues of~$H$ are just shifted by $\varepsilon^{-2} E_1$
with respect to those of~$H_{\mathrm{ren}}$, while the eigenfunctions coincide.

Let us remark at this point that, based on our estimates,
it is possible to give 
explicit threshold values for the smallness of~$\varepsilon$ 
(in terms of the geometry of~$\Sigma$, the magnetic field, 
and the isolation distance of~$\mu$).
However, we have preferred this more concise presentation 
for the sake of readability of the paper.

\subsection*{Acknowledgment}
Support by the Institute Mittag-Leffler (Djursholm, Sweden), 
where this paper was being prepared, is gratefully acknowledged. 
The work has been partially supported
by the projects RVO61389005 and RVO68407700;
and the grants No.\ P203/11/0701 and No.\ 13-11058S of the Czech Science Foundation (GA\v{C}R).

%
{\small

\providecommand{\bysame}{\leavevmode\hbox to3em{\hrulefill}\thinspace}
\providecommand{\MR}{\relax\ifhmode\unskip\space\fi MR }
\providecommand{\MRhref}[2]{%
  \href{http://www.ams.org/mathscinet-getitem?mr=#1}{#2}
}
\providecommand{\href}[2]{#2}

}

\end{document}